\documentstyle[12pt]{ioplppt}

\begin{document}
\jl{3}

\title{Magnetic excitation spectra of $\rm NaV_2O_5$}

\author{K Ladavac\dag, M J Konstantinovi\' c\dag, A Beli\' c\dag, Z V Popovi\' c\dag, A N Vasil'ev\ddag, M Isobe\S and Y Ueda\S}

\address{\dag\ Institute of Physics, P.O.Box 57, 11001 Belgrade, Yugoslavia}

\address{\ddag\ Low Temperature Physics Department, Moscow State University, 119899 Moscow, Russia}

\address{\S\ Institute for Solid State Physics, The University of Tokyo, 7-22-1 Roppongi, Minato-ku, Tokyo 106, Japan}

\begin{abstract}
We present low temperature Raman spectra of $\rm NaV_2O_5$ single crystal. The assignation of the observed modes is done on the basis of Monte Carlo calculation of excited states of Heisenberg dimerized Hamiltonian, without frustration. We found a good agreement between measured and calculated magnetic excitations for $\delta=0.048$ and $J=452{\rm\: K}$. The singlet-triplet and singlet-singlet excitations as well as two-magnon mode are observed to be at $66$, $127$ and $132{\rm\: cm^{-1}}$.
\end{abstract}


\section{Introduction}

It has been recently shown \cite{1} that sodium vanadium oxide is the second  example of the inorganic spin-Peierls compound following $\rm CuGeO_3$ \cite{2}. Similarly to $\rm CuGeO_3$ the strong decrease of the susceptibility $\chi(T)$ in $\alpha'-\rm NaV_2O_5$ is found below  $T\cong34\rm\: K$ as a consequence of the opening of a spin-gap. The spin-Peierls transition is one of the most interesting phenomena observed in low-dimensional quantum spin systems. It occurs in crystals containing linear chains of half-integer spin coupled by antiferromagnetic exchange interaction. The spin-Peierls transition is analogue of the Peierls transition in metal and results in dimerisation of the lattice induced by magnetic fluctuations. The lattice dimerisation and the existence of the spin-gap are also confirmed by x-ray, inelastic neutron scattering \cite{3}, NMR \cite{4} and ESR \cite{5} measurements. Using these techniques \cite{3,4,5}, the spin gap was estimated to be in the range from $85$ to $98\rm\: K$. The previously published infrared \cite{6,7} and Raman \cite{8,9} spectra show the appearance of new modes below the transition temperature, but nothing was said about their origin.

Based on an early measurement of the $\rm NaV_2O_5$ crystal structure \cite{10}, it has been shown that this oxide, grown in single crystalline form under ambient conditions, has an orthorhombic unit cell  with parameters $a=1.1318 {\rm\: nm}$, $b=0.3611 {\rm\: nm}$, $c= 0.4797{\rm\: nm}$, $Z=2$ and the space group  $P2_1mn$ ($D_{2v}^7$). Such a crystalline structure assumes two kinds of vanadium chains along the b-axis. One is magnetic $\rm V^{4+}$ ($S=1/2$) and the other one is a nonmagnetic $\rm V^{5+}$ ($S=0$ chains). These chains are aligned parallel to each other by sharing an edge of adjacent ${\rm VO_5}$ pyramids, thus forming layers in $ab$ plane. ${\rm Na}$ atoms are situated between layers as intercalates. This magnetic structure is expected to be truly one-dimensional since the chains caring the spin are separated from each other by nonmagnetic chains along $b$-axis. The problem with this structure is missing physical argument for such a strong charge modulation. The most recent structural analysis \cite{11} shows that all vanadium atoms are equivalent and in mixed valent state. Such a crystal structure is also predicted from polarized Raman scattering measurements \cite{12,13}.  Even without charge modulation, due to strong  electron correlations, it is possible to have formation of antiferromagnetic Heisenberg chains \cite{14}. It is also possible that an induced zigzag type of charge ordering along the ladders of V-ions is formed \cite{14a}. However, in this paper we consider only 1-D antiferromagnetic Heisenberg model as an explanation for the magnetic behaviour of this compound. This is based on the magnetic susceptibility measurements of $\rm NaV_2O_5$  powder \cite{1} which shows a broad maximum at $350 \rm\: K$, followed by rapid decrease below $34 \rm\: K$. Such a susceptibility behaviour can be well described as linear Heisenberg antiferromagnet in the high temperature phase and by Bulaevskii theory \cite{8} in dimerized phase, indicating that there is no frustration in this material.

In this work we present low temperature Raman spectra of $\rm NaV_2O_5$  single crystal. The assignation of the observed modes is done on the basis of Monte Carlo calculation of excited states of Heisenberg dimerized Hamiltonian, without frustration. We found a good agreement between measured and calculated magnetic excitations for $\delta=0.048$ and $J=452 \rm\: K$. The singlet-triplet and singlet-singlet excitations as well as two-magnon mode are observed to be at $66$, $127$ and $132{\rm\: cm^{-1}}$.

\section{Experiment}

The single crystals of $\alpha'-\rm NaV_2O_5$ with a size of approximately $1\times4\times0.5 {\rm\: mm^3}$ are obtained as described in \cite{15}. The spectral lines of an argon-ion laser are used as an excitation source. The light beam was focused on the sample surface using cylindrical lens with an average power of $100 {\rm\: mW}$. The scattered light was collected with an objective, aperture 1:1.4. The monohromator used was a Jobin Ivon model U1000, equipped with a Pelletier-effect-cooled RCA 31034 photomultiplier with a conventional photon counting system. The temperature was controlled with closed-cycle helium cryostat (Leybold).

Numerical calculation were carried out on a 4-processor Origin 2000 computer, courtesy of Institute of Physics Computer Facilities (IPCF).

\section{ Raman spectra}

The polarized Raman spectra measured with $514.5 {\rm\: nm}$ laser line at temperature of $15 \rm\: K$ are presented in \Fref{slika1}. The new modes found at temperatures below transition temperature $T=34 \rm\: K$ are denoted in the graph. The lowest excitation is observed at $66 {\rm\: cm^{-1}}$, with the largest intensity in $ab$ polarized spectra. This mode is observed in $bb$ polarization as well. The continuum of magnetic excitations starts at $132 {\rm\: cm^{-1}}$ which is exactly two times $66 {\rm\: cm^{-1}}$ energy and is observed only in $ab$ configuration. The $106 {\rm\: cm^{-1}}$ mode is found to be largest in $bb$ spectrum.
The other modes in the spectra are high temperature phonons assigned on the basis of valence shell model calculation \cite{13}. The broad structure present in $aa$ spectra is assigned as electronic continuum which appears due to $d-d$ electronic transitions of $\rm V^{4+}$ ions \cite{16}.

To further analyse the origin of new modes we measured the Raman spectra at different temperatures from $15$ to $30 \rm\: K$, \Fref{slika2}. We found that $66 {\rm\: cm^{-1}}$ mode exhibits the shift towards lower energy of $\sim 3 {\rm\: cm^{-1}}$ by increasing temperature. Besides the $132 {\rm\: cm^{-1}}$ mode that shows similar shift, we observed additional structure at $127 {\rm\: cm^{-1}}$ in the $T=15 \rm\: K$ spectrum, denoted with an arrow. To resolve these two structures we performed higher resolution measurements (resolution$<1.5 {\rm\: cm^{-1}}$), see inset \Fref{slika2}. By increasing the temperature this part of the spectrum rapidly looses intensity and the difference between these two modes is no longer observable.

Moreover, the existence of the continuum is confirmed by observing the Fano-type asymmetry of the $177 {\rm\: cm^{-1}}$ phonon with energy overlapping with that of the continuum. The asymmetry of this mode is becoming less pronounced by increasing temperature and finally becoming complete Lorenzian in the $T=30\rm\: K$ spectrum.

Other modes in the spectra do not show frequency shift and we assigne them as phonons becoming active due to doubling of the unit cell.

\section{Monte Carlo calculation}

Here we present the results of the numerical calculation based on Green's Function Monte Carlo technique (GFMC) \cite{17}. We assumed that $\rm NaV_2O_5$   can be modelled as 1\nobreak-\nobreak D Heisenberg antiferomagnet with modulated NN coupling and no frustration. The relevance of this model comes from susceptibility measurements \cite{8} were it was shown that low temperature behaviour can be well described with the Hamiltonian:

\begin{equation}
{\cal H}=J\sum_i (1+(-1)^i\delta)\vec S_i\vec S_{i+1}
\end{equation}
where $\vec S_i$ are spin-1/2 operators. Further, we assumed the periodic boundary conditions so that only rings with even numbers of spins were studied. We have calculated energies of the ground, the first excited triplet ($S=1$) and quintuplet ($S=2$) states for rings with $N=6, 10, 20, 100, and 200$ spins. All these energies are calculated exactly with no approximation. The resulting triplet and quintuplet gaps, $\Delta_T$ and $\Delta_Q$, are shown in \Fref{slika3} for $\delta=0.05$. The extrapolation to $N\to\infty$ is done using a $\Delta_{T,Q}(N)=\Delta_{T,Q}+(A/N+B/N^2+C/N^3)exp(-N/N_0)$ function. Similar simpler functions were used in \cite{18} to extrapolate gaps obtained by Lanczos diagonalization for systems up to $N=28$. Here, we have chosen more complicated ansatz to fit points for smaller $N$ also. The estimated error for the extrapolated values of triplet and quintuplet gaps are less then 1\%. The size of symbols in figures do not represent this error.
 Further details of extrapolation procedure can be found in \cite{19}.

The extrapolated results for all $\delta$ considered are given in \Fref{slika4}. It is found that the singlet-triplet gap can be excellently fitted by the simple formula $\Delta_T/J=2\delta^{3/4}$, $0.01<\delta<1$.
This was also found in the recent study \cite{20}. In the limit $\delta\to1$ this formula coincides with leading three orders of the perturbation theory, \cite{21}. The $\delta<0.01$ region was of no concern to us, and we do not know were the crossover to the known \cite{22,23} critical behaviour $\Delta_T/J\sim\delta^{2/3}/\sqrt{\vert\ln{\delta}\vert}$ occurs.

The singlet-quintuplet gap can not be fitted in such a simple fashion. We have found that it lies above the double singlet-triplet gap value, at the energy $\Delta_Q=2\Delta_T+\epsilon$. This is consistent with the RPA study \cite{24} were this mode was identified as an anti-bound two-magnon state. The anti-boundedness $\epsilon$ decreases towards zero by increasing $\delta$ to 1. According to the estimated error of our calculation the existence of a quintuplet state at energy differenet then 2$\Delta_T$ is established for $\delta\le 0.1$, see inset \Fref{slika4}.

\section{Discussion}

To begin discussion, first we compare our GFMC calculation with other numerical results. The ground state energy is in excellent agreement with Bethe ansatz solution \cite{25} in the limit $\delta\to0$ and with the results for isolated dimmers when $\delta\to1$ \cite{21}. Moreover, our values for the triplet gap coincide precisely with the exact diagonalization results \cite{26}. The most important point, which is one of our main numerical results, is exact calculation for the lowest lying quintuplet state, $S=2$. As allready mentioned above, this mode at energy $\Delta_Q=2\Delta_T+\epsilon$ was identified as an anti-bound two-magnon state \cite{24}. According to that study one should also expect singlet, two-magnon bound state, at the energy $\Delta_S=2\Delta_T-2\epsilon$. Such an excited singlet state can not be calculated exactly using GFMC but it can be estimated using the above formula.

The low temperature susceptibility measurements \cite{8} can be well described with Bulaevskii theory, indicating the spin-Peierls transition at 34 K. The fit gave value of the exchange integral $J$ and dimerization $\delta$ to be $J=441 \rm\: K$ and $\delta=0.048$. Using slightly different parameters $J=452\rm\: K$ and $\delta=0.048$ we calculated singlet-triplet and singlet-quintuplet transition energies to be $\Delta_T=66 {\rm\: cm^{-1}}$ and $\Delta_Q=134.5 {\rm\: cm^{-1}}$. These values are obtained by a linear interpolation between calculated values for $\delta=0.04$ and $\delta=0.05$.
The value for singlet-singlet transition $\Delta_S=127 {\rm\: cm^{-1}}$ is obtained using above estimate. This value is in good agreement with exact diagonalization study which gives singlet to triplet gap ratio of $\sim1.9$ \cite{18}. In the Raman spectra the $66 {\rm\: cm^{-1}}$ feature coincides with our numerical result for the triplet gap and it is possible that this mode comes from one-magnon scattering process. This value is in agreement with other measurements of the spin-gap. Moreover, this mode shifts towards lower energies by increasing temperature and by decreasing sodium concentration \cite{27}. It was proved that sodium deficiency strongly suppresses the SP transition \cite{28}, which is an additional proof for the magnetic origin of this mode. Still, behaviour of this mode in magnetic field remains to be studied. Some preliminary results show no change of this mode in the magnetic fields up to $7 \rm\: T$ \cite{29}.

The two-magnon mode is found to be at $132 {\rm\: cm^{-1}}$ which is exactly twice the gap energy. It appears as a well-defined peak due to the one-dimensionality of the magnetic interaction. Then the structure at $127 {\rm\: cm^{-1}}$ can be described as singlet-singlet excitation with boundness $2\epsilon=5{\rm\: cm^{-1}}$, which is in agreement with our calculation.

It  is interesting to compare our results with that of $\rm CuGeO_3$ which is the first inorganic material  exhibiting the spin-Peierls transition. The magnetic mode at $30 {\rm\: cm^{-1}}$ observed in $\rm CuGeO_3$  \cite{30} is described as excitation from the ground state to singlet state with singlet to triplet gap ratio of $1.76$. Such a ratio is a consequence of frustration, which is quite large in this material. From our Raman spectra we found that in $\rm NaV_2O_5$ this ratio is around $1.95$, which is consistent with assumption that frustration is negligible in this material. Moreover, the broad magnon band, centred at about twice the maximum magnon energy is not observed in our Raman spectra which again proves existence of no frustration in $\rm NaV_2O_5$ \cite{18}.

The appearance of magnetic scattering in $ab$ polarized configuration can not be explained using exchange scattering mechanism of Fleury and Loudon \cite{31}, since this theory predicts the magnetic scattering in configuration parallel to the chains, ie. in $bb$ direction. Our previous study of Raman spectra \cite{16}, using different laser line energies shows existence of resonance effects which may influence selection rules for magnetic scattering. It is also shown that $66 {\rm\: cm^{-1}}$ feature, in the spectra measured with $647.1 {\rm\: nm}$ laser line consists of two modes separated by $\sim 2 {\rm\: cm^{-1}}$. Still, the mechanism for the apearance of the  second mode is not yet established. It was proposed that this mode, as well as, the $106 {\rm\: cm^{-1}}$ comes from crystal field excitations of $\rm V^{4+}$ ions. 

Finally, we conclude that magnetic excitations in the Raman spectra of $\rm NaV_2O_5$ can be well described using  GFMC calculation of excited states of dimerized 1-D Heisenberg antiferomagnet with $J=452\rm\: K$, $\delta=0.048$ and no frustration.

\ack

This work is supported by the Serbian Ministry of Science and Technology under Projects No. 01E10M1, 01E15 and 01E09.

\Figures

\Figure{\label{slika1}
The polarized Raman spectra of $\rm NaV_2O_5$ at $T=15 \rm\: K$. The modes that apear at temperatures below $T=34\rm\: K$ are denoted in the graph.
}

\Figure{\label{slika2}
The $ab$ polarized Raman spectra as a function of temperature. The arrows indicate position of magnetic modes, see text. Vertical dashed lines are placed to clarify the energy shift. Inset: The $15\rm\: K$ $ab$ Raman spectra measured with increased resolution.
}

\Figure{\label{slika3} 
The 
triplet and 
quintuplet gaps versus $1/N$, for $\delta=0.05$. The 
lines are fitting functions defined in the text.}

\Figure{\label{slika4}
The  extrapolated values, $N\to\infty$, for 
triplet and 
quintuplet gaps as functions of $\delta$. The full
line is a function $\Delta/J=2\delta^{3/4}$. The dashed
line is a function $\Delta/J=4\delta^{3/4}$, see text.
Inset: the gap values with error bars for some $\delta$.}

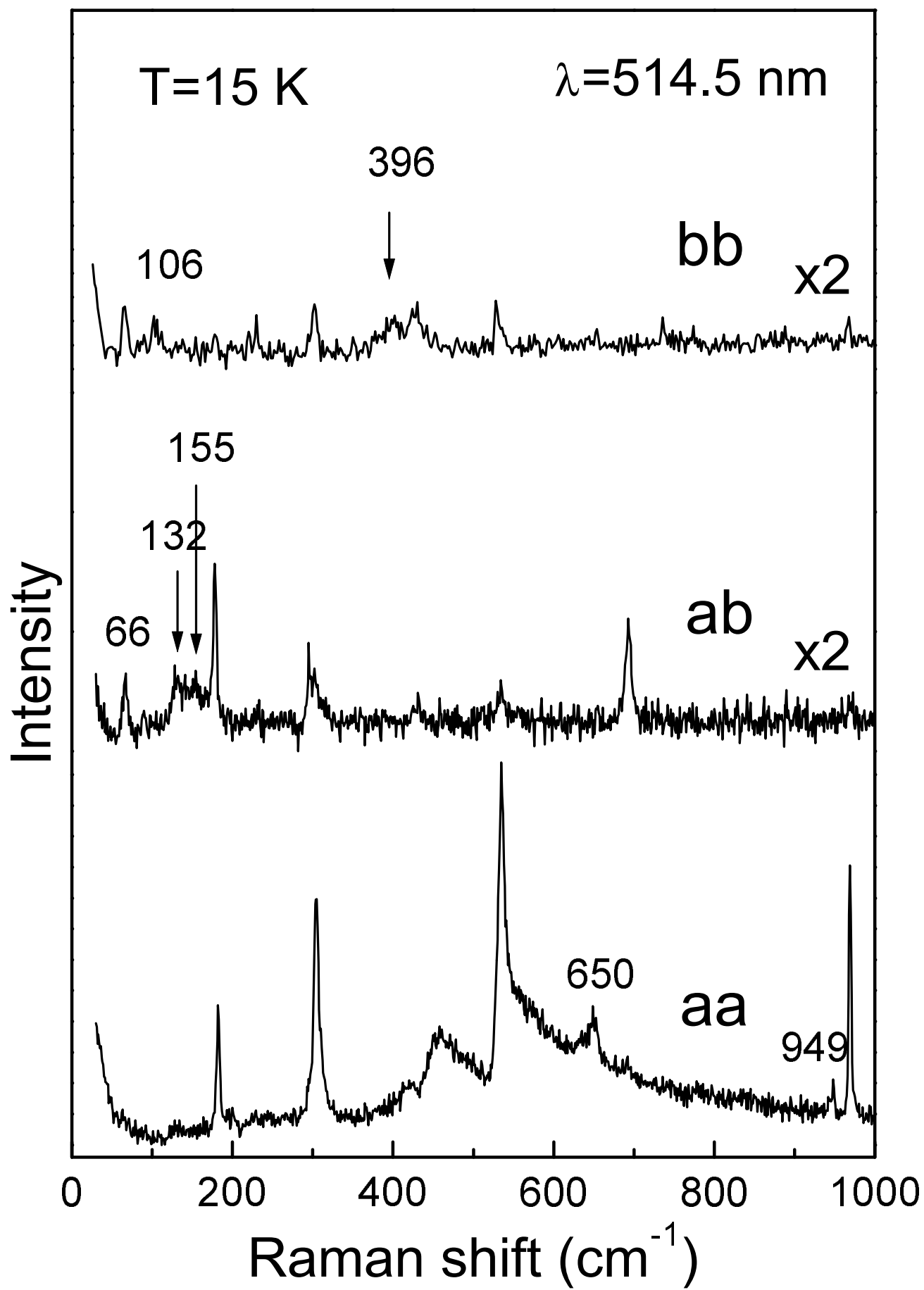
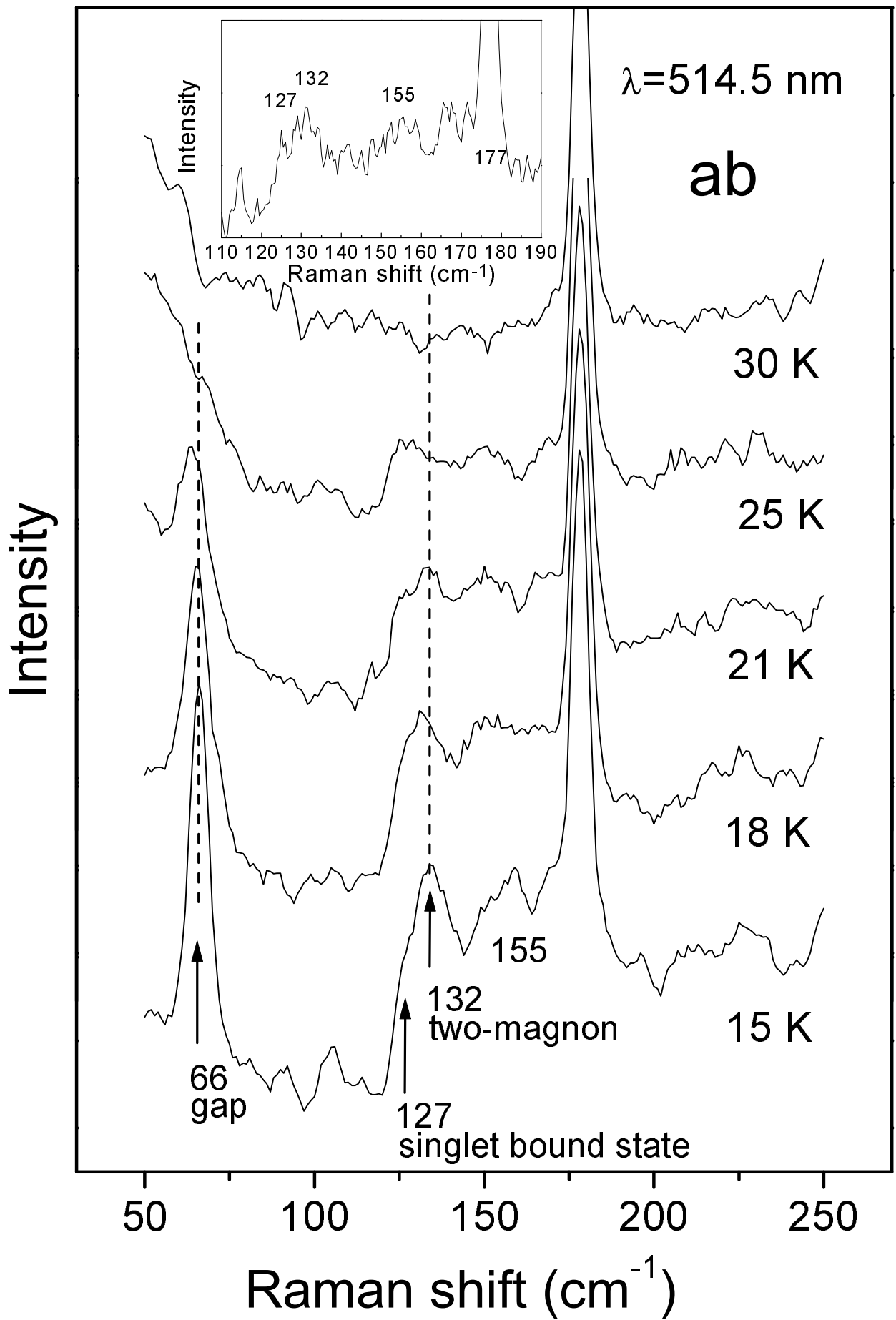
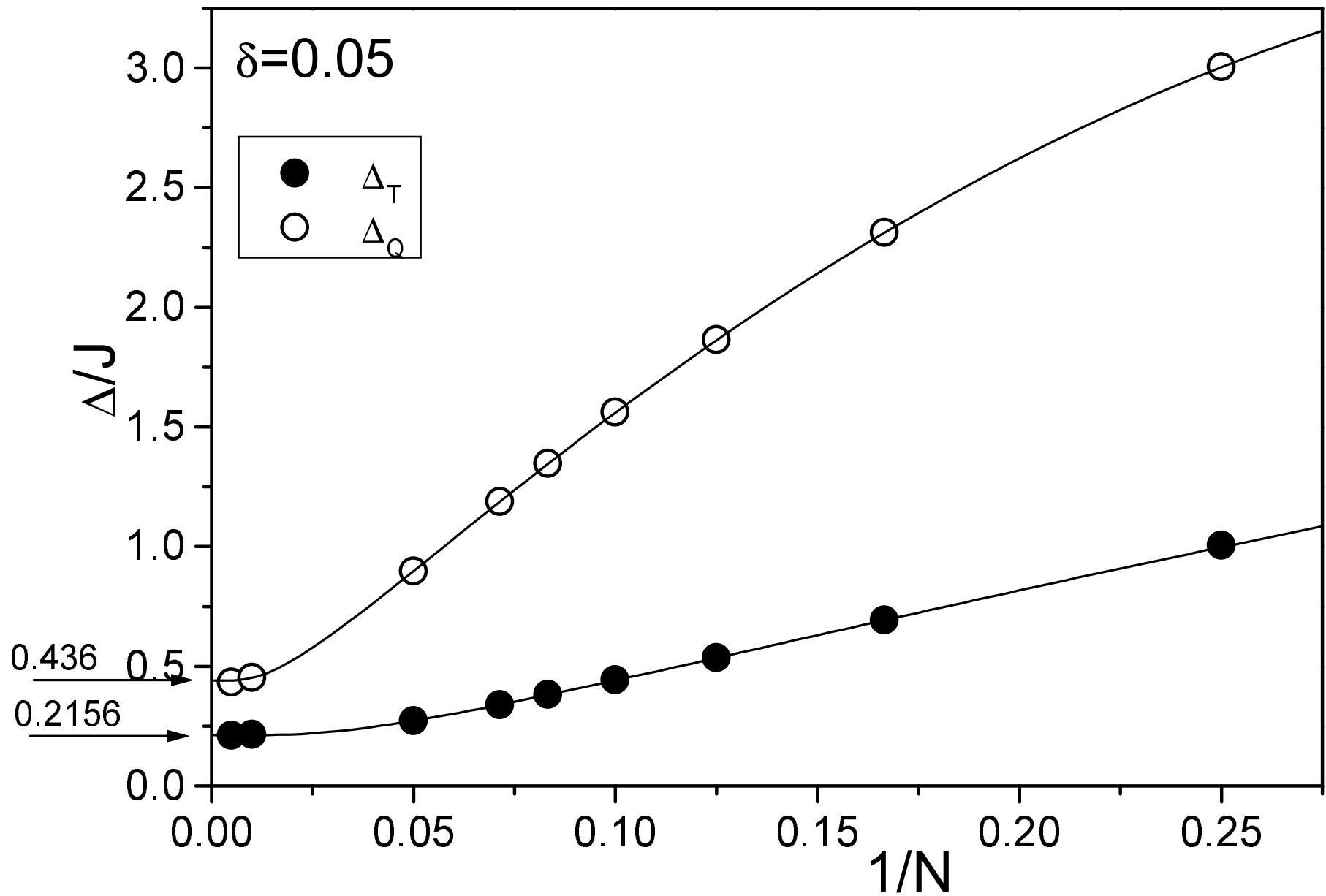
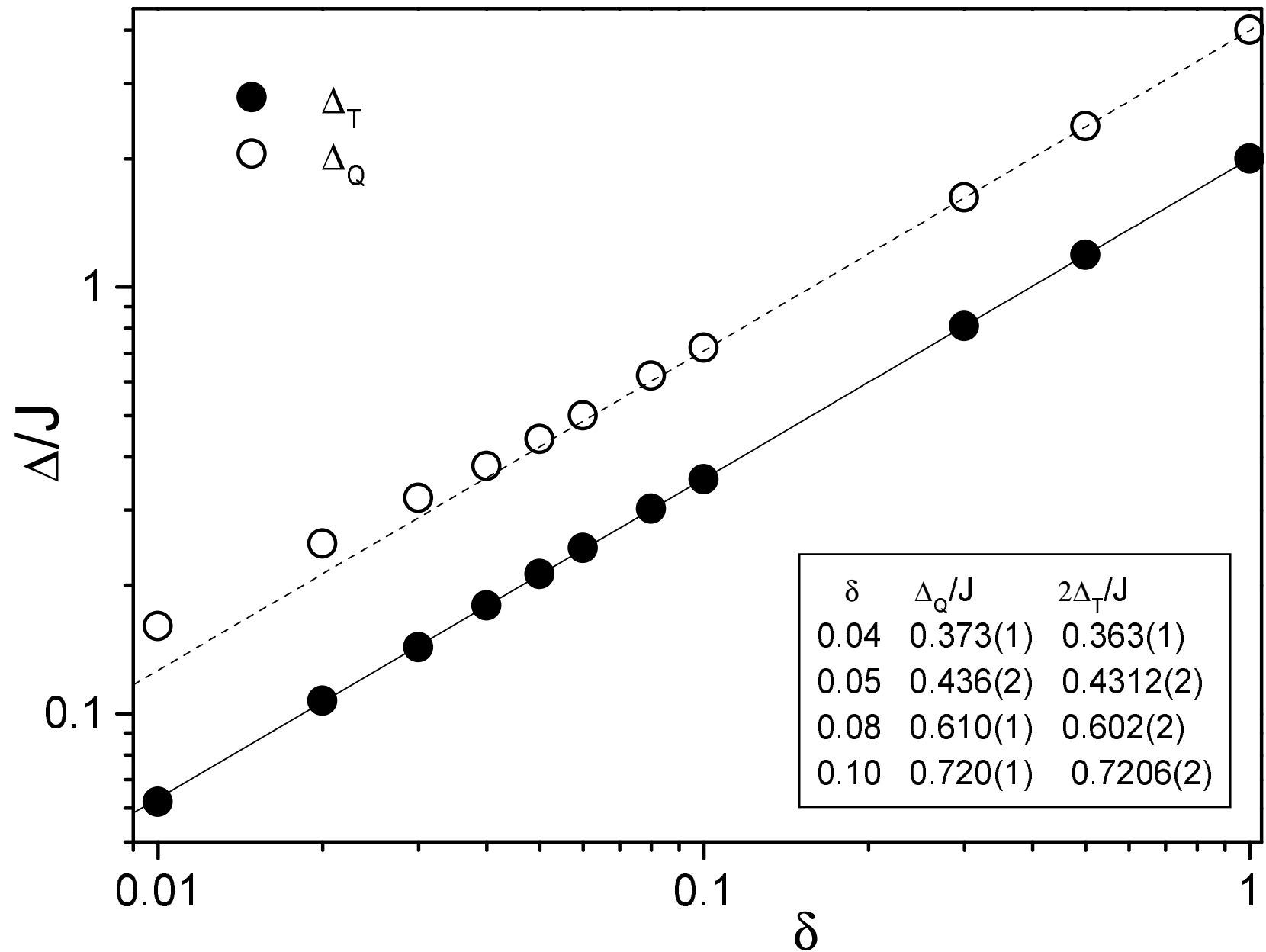
\end{document}